\documentclass[twocolumn,superscriptaddress,amsmath,amstext,amssymb,floatfix,preprintnumbers]{revtex4-1}
\usepackage{slashed}
\usepackage{graphicx}
\usepackage{xcolor}
\usepackage[pdftex,final]{hyperref}

\newcommand{\TEMT}{ \Tud{\rho}{\rho} }
\newcommand{\Tud}[2]{T^{#1}_{\;\; #2}}
\newcommand{\psib}{\bar{\psi}}
\newcommand{\lam}[1]{\lambda_{#1}}
\DeclareMathOperator{\Tr}{Tr}
\newcommand{\vev}[1]{\langle #1 \rangle} 
\newcommand{\state}[1]{|#1\rangle}
\newcommand{\matel}[3]{\langle #1|#2|#3\rangle}
\newcommand{\ORD}{{\cal O}}
\newcommand{\Op}{{\cal O}}
\newcommand{\de}{\delta}
\newcommand{\De}{\Delta}
\newcommand{\la}{\lambda}
\newcommand{\Oqq}{\bar \psi \psi}
\newcommand{\Fdec}{F_D}
\newcommand{\vphiz}{\varphi_0}
\newcommand{\vphi}{\varphi}
\newcommand{\lamcrit}{\lambda_3^{\rm crit}}

\begin{document}

\title{Dilaton Physics from Asymptotic Freedom}
\preprint{CERN-TH-2024-098}

\author{Charlie Cresswell-Hogg}
\email{c.cresswell-hogg@sussex.ac.uk}
\affiliation{Department of Physics and Astronomy, University of Sussex, Brighton, BN1 9QH, U.K.}

\author{Daniel F.~Litim}
\email{d.litim@sussex.ac.uk}
\affiliation{Department of Physics and Astronomy, University of Sussex, Brighton, BN1 9QH, U.K.}
\affiliation{Theoretical Physics Department, CERN, 1211 Geneva 23, Switzerland}

\author{Roman Zwicky}
\email{roman.zwicky@ed.ac.uk}
\affiliation{Higgs Centre for Theoretical Physics, School of Physics and Astronomy, University of Edinburgh, Edinburgh EH9 3JZ, Scotland}

\begin{abstract}
The dilaton is investigated from first principles in an asymptotically free Gross-Neveu-Yukawa theory in three dimensions. In the limit of many fermion flavours, the theory features a finite line of strongly interacting fixed points with continuous quantum phase transitions and a massless Goldstone boson, the dilaton, following  spontaneous  scale symmetry breaking at its endpoint. Interestingly, we find that the emergence of a vacuum expectation value and a dilaton can be understood as a double-scaling limit. Exploiting the scalar two-point function, we identify the dilaton in the spectrum, compute its decay constant, and obtain universal expressions for the induced dilaton mass in terms of small perturbations. Consistency of findings with soft dilaton theorems is equally established. Implications for spontaneously broken conformal theories are indicated.
\end{abstract}

\maketitle

{\it Introduction.---} 
Scale symmetry is an important concept in quantum and statistical physics.
It arises at fixed points of the renormalisation group, often alongside full conformal symmetry, and
implies that theories are massless with correlation functions   given  in terms of universal numbers, the scaling dimensions. 
New phenomena arise when scale symmetry is broken spontaneously  \cite{Isham:1970xz,Isham:1970gz,Zumino:1970tu,Leung:1989hw,Zwicky:2023krx} 
leading to a massless Goldstone boson -- the dilaton -- and  the appearance of a new mass scale 
that is not determined by  microscopic parameters of the theory \cite{Bardeen:1983rv,Bardeen:1983st,David:1984we,David:1985zz,Eyal:1996da,Omid:2016jve,Marchais:2017jqc,Litim:2018pxe,Cresswell-Hogg:2022lgg,Cresswell-Hogg:2022lez,Bardeen:1984dx,Litim:2011bf,Heilmann:2012yf,Aharony:2012ns,Bardeen:2014paa,Moshe:2014bja,Cresswell-Hogg:2023hdg,Semenoff:2024prf,Cresswell-Hogg:2024pxd,Semenoff:2024jqf}. 
The dilaton offers a rich phenomenology  \cite{Ellis:1970yd,Ellis:1971sa,Crewther:1970gx,Crewther:1971bt,Zwicky:2023fay} with applications in QCD \cite{Crewther:2012wd,Crewther:2015dpa,DelDebbio:2021xwu,Zwicky:2023bzk,Shifman:2023jqn}, lattice field theory \cite{Hasenfratz:2024fad,LatticeStrongDynamics:2023bqp,Hasenfratz:2020ess,Fodor:2017gtj,Chiu:2018edw,LatticeStrongDynamics:2018hun,LatKMI:2016xxi,DelDebbio:2015byq,LatKMI:2014xoh}, dense nuclear interactions \cite{Brown:1991kk,Ma:2019ery,Rho:2021zwm,Shao:2024qny}, models of particle physics \cite{Dietrich:2005jn,Goldberger:2007zk,Bellazzini:2012vz,Matsuzaki:2012xx,Cata:2018wzl,Bruggisser:2022rdm,Appelquist:2024koa},
cosmology \cite{Wetterich:1987fm,Shaposhnikov:2008xb,Shaposhnikov:2008xi,Csaki:2014bua,Cacciapaglia:2023kat}, and holography \cite{Pomarol:2019aae,Faedo:2024zib}. 

Solvable models with  scale symmetry breaking  and a dilaton are hard to come by. 
In three dimensions, however,  and in the limit of many matter fields, a range of models is available including  scalar 
\cite{Bardeen:1983rv,Bardeen:1983st,David:1984we,David:1985zz,Eyal:1996da,Omid:2016jve,Marchais:2017jqc,Litim:2018pxe}, 
 fermionic  \cite{Cresswell-Hogg:2022lgg,Cresswell-Hogg:2022lez}, 
supersymmetric  \cite{Bardeen:1984dx,Litim:2011bf,Heilmann:2012yf}, 
Chern-Simons-matter~\cite{Aharony:2012ns,Bardeen:2014paa,Moshe:2014bja}, 
and Gross-Neveu--Yukawa   \cite{Cresswell-Hogg:2023hdg,Semenoff:2024prf,Cresswell-Hogg:2024pxd,Semenoff:2024jqf} theories. 
As such, they provide useful opportunities to understand dilaton physics directly from a microscopic theory. 

In this Letter we investigate the Gross-Neveu--Yukawa theory of   \cite{Cresswell-Hogg:2023hdg} as a template for dilaton physics. 
The theory features asymptotic freedom and a  line of  low-energy fixed points with continuous quantum phase transitions. We explain where and why  spontaneous scale symmetry  breaking occurs and demonstrate how this phenomenon can be understood as a double-scaling limit. We   identify the  dilaton in correlation functions and compute its decay constant. Most notably, we  find universal expressions for the  dilaton mass induced by small perturbations that break scale symmetry explicitly, and establish consistency with soft dilaton theorems \cite{Zwicky:2023krx}. Further implications, including for conformal field theories (CFTs) in the presence of a vacuum expectation value, are  provided.

{\it Yukawa theory with a conformal window.---} 
To set the stage, we consider $N$  two-component Dirac fermion $\psi_a$ coupled to a real scalar field $\phi$ in three euclidean dimensions  \cite{Cresswell-Hogg:2023hdg}. The fundamental  action  contains  a cubic scalar self-coupling $\bar\lambda_3$ and a Yukawa coupling $H$, 
\begin{equation}
\label{eq:SGNY}
S=\int _x \left[\psib_a   \slashed{\partial}  \psi_a \! + \tfrac12 ( \partial \phi )^2 + H  \phi  \mkern2mu \psib_a \psi_a \! + \tfrac1{3!}\bar \lambda_3\phi ^3 \right]\,. 
\end{equation}
Mass terms will be added later, and other  renormalisable interactions play no role for what follows. This 
Gross-Neveu--Yukawa theory is perturbatively renormalisable and asymptotically free in the ultraviolet (UV) where all fundamental interactions are relevant. 
Under the renormalisation group flow, and  in the large-$N$ limit, the cubic  coupling $\bar\lambda_3$ becomes exactly marginal 
and leads to a line of strongly interacting fixed points in the  infrared (IR), characterised by the  parameter $\lam{3} = \bar\lambda_3 / H^3$. 
The RG-invariant  scale  $H^2$  controls the crossover from asymptotic freedom to  conformality \cite{Cresswell-Hogg:2023hdg}. 

\begin{figure}[!t]
\includegraphics[width=.9\linewidth]{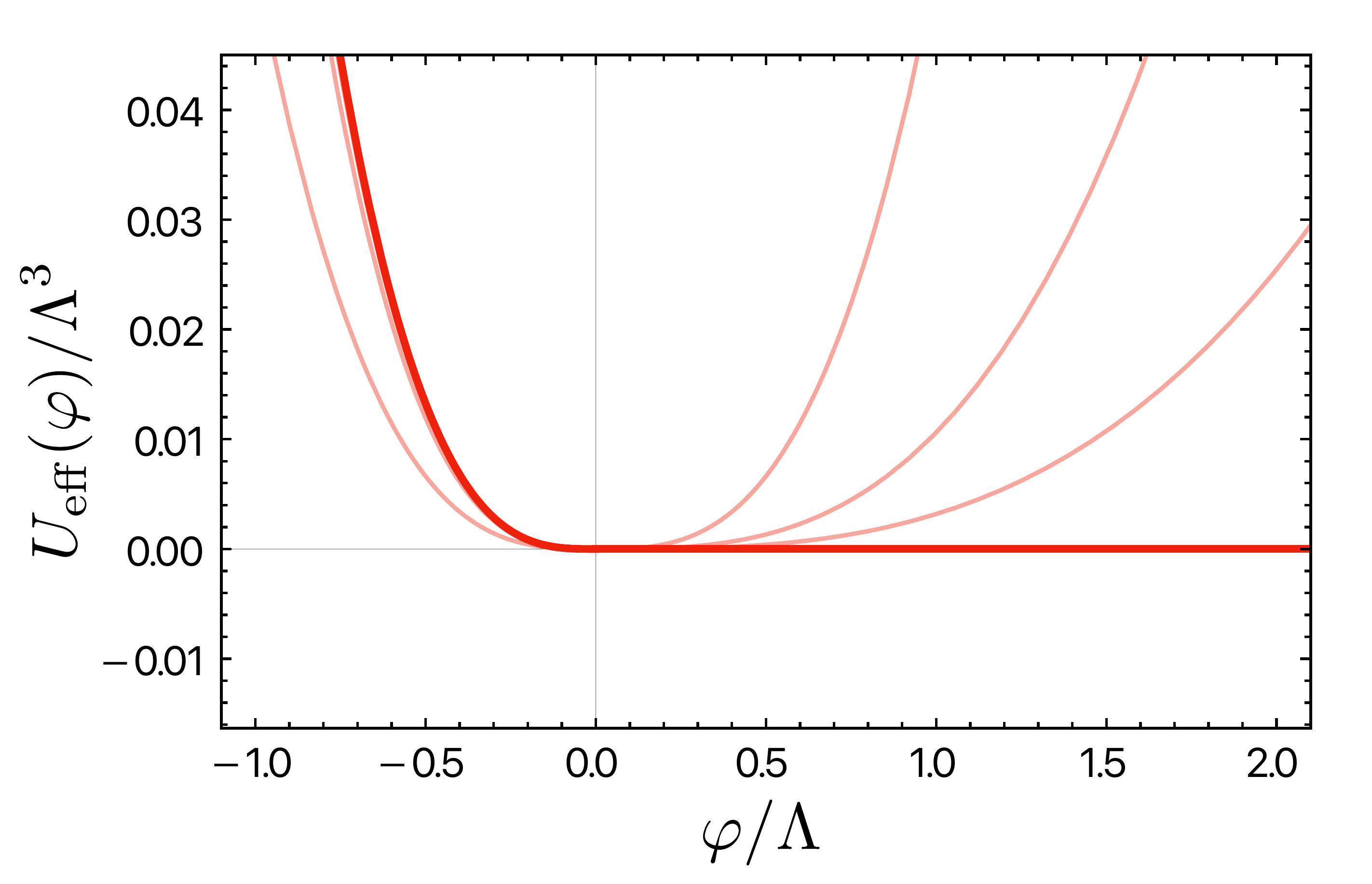}
\caption{Quantum critical potentials in the conformal window for increasing $0\le |\lambda_3|\le \lambda^{\rm crit}_3$ (light red), approaching the endpoint with broken scale symmetry (thick red), and $\Lambda$ being an arbitrary scale.}
\label{fig:UeffCrit}
\end{figure}

The  parameter regions of the theory \eqref{eq:SGNY}  with IR conformal scaling  can be identified from the  quantum effective potential. Given that 
field monomials $\phi^n$ develop non-canonical scaling dimensions  $\De_{\phi^n} = n$ at  fixed points, it is natural to  work with the rescaled  field $\vphi = H \phi$ \cite{Cresswell-Hogg:2023hdg}.  Quantum critical potentials of the theory \eqref{eq:SGNY} then take the simple form 
\begin{equation}
\label{eq:Ueff}
U_{\rm eff} ( \vphi ) = \frac{1}{3!} \lam{3} \, \vphi^3 + \frac{1}{3!} \lamcrit | \vphi |^3 \,.
\end{equation}
Fluctuations have generated  the non-analytic term $| \vphi |^3$ with    universal   coupling $\lamcrit=\tfrac1\pi$.
We  observe a  {\it finite} conformal window of IR  fixed points 
with a stable  ground state, limited by the cubic coupling taking values within  $|\lam{3} | \leq \lamcrit$  \cite{Cresswell-Hogg:2023hdg}. 
We further note that  under $\varphi\to-\varphi$,  the conformal window is mapped onto itself via $\lam{3}\to -\lam{3}$.
 The critical potentials \eqref{eq:Ueff} are shown in Fig.~\ref{fig:UeffCrit} for various $\lam{3}<0$.  The potential becomes half-sidedly flat when $\lam{3}$ approaches the critical coupling strength $| \lam{3} | \to \lamcrit$. Exactly at the endpoint, we observe a moduli space of degenerate vacua. Hence, even though the vacuum expectation value of a scalar field $\langle\varphi\rangle\equiv\varphi_0$ vanishes  identically {\it inside}  the conformal window, as it must for a conformal theory, 
  $\varphi_0\neq 0$   becomes a new fundamental parameter at the  endpoint. 
It follows that the fermions become massive $m = \varphi_0$, while the scalar remains massless.

\begin{figure}[b]
\includegraphics[width=.9\linewidth]{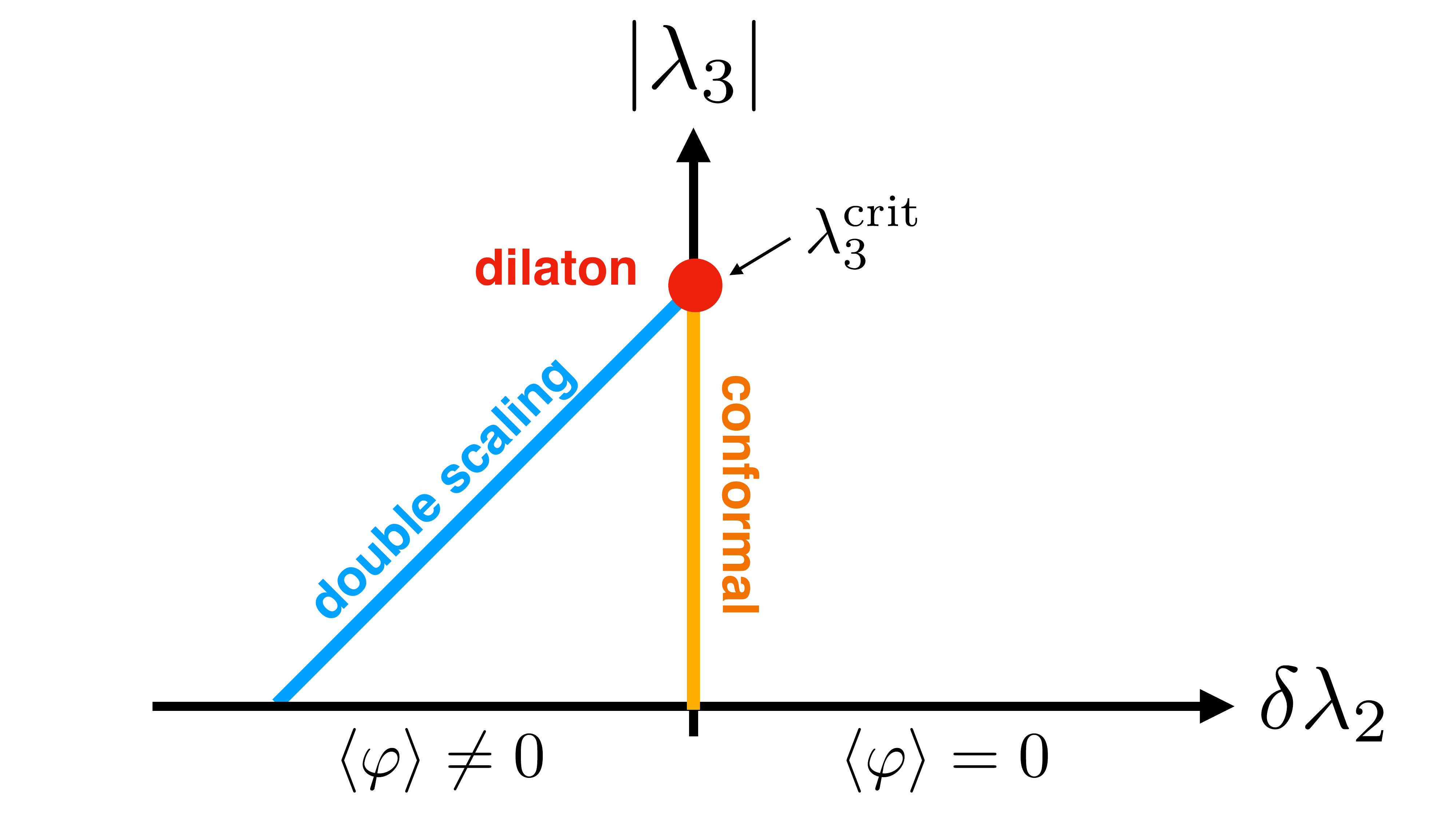}
\caption{Phase structure of the Gross-Neveu--Yukawa theory in the presence of a scalar mass perturbation $\delta\lambda_2$, showing the conformal window of  critical points (orange line), its endpoint with broken scale symmetry and a dilaton (red dot), and the double-scaling line \eqref{eq:DSlimit} (blue line).}
\label{fig:doubleScaling}
\end{figure}

{\it Quantum phase transitions and double-scaling limit.---}
The  quantum critical behaviour of the theory \eqref{eq:SGNY}  is further appreciated  by adding a small scalar mass term to dial through the  phase transition of  \eqref{eq:Ueff} at vanishing mass.
Without loss of generality we consider $\lam{3} < 0$, and  parametrise the distance from the endpoint of the conformal window by $\de\lam{3} = \lam{3} + \lamcrit > 0$. Adding a  scalar mass term in  the theory \eqref{eq:SGNY}  leads to a  mass term $\propto \de\lam{2}$ in the quantum effective action  \eqref{eq:Ueff}, giving
\begin{equation}
\label{eq:Ueff23}
U_{\rm eff} ( \vphi ) = \frac12 \de\lam{2} \, \vphi^2 + \frac{1}{3!} \de\lam{3} \, \vphi^3 + \frac{1}{3!} \lamcrit ( | \vphi |^3 - \vphi^3 ) \,.
\end{equation}
We first consider the  critical behaviour away from the endpoint by fixing $\delta\lam{3}>0$. For positive $\de\lam{2}$, the global minimum of \eqref{eq:Ueff23} corresponds 
to a vanishing expectation value $\vev{\vphi} = 0$. 
This ``disordered'' phase is characterised by a massive scalar field and massless  fermions  \cite{Cresswell-Hogg:2024pxd}.
For negative $\de\lam{2}$, instead, the global minimum   is located at $\varphi_0>0$ with
\begin{equation}
\label{eq:VEV}
\vphiz = -2\frac{\de\lam{2}}{\de\lam{3}} \,,
\end{equation}
and the theory enters an ``ordered'' phase where 
 fermions acquire a mass due to the scalar condensate while the scalar field remains massive. 
For any $\delta\lam{3}>0$ in the conformal window, the two phases are separated by a second order quantum phase transition  at $\delta\lam{2}=0$  where all fields are massless, illustrated   in Fig.~\ref{fig:doubleScaling} (orange line). 
The transition does not break any global  symmetries, except parity symmetry if $\lam{3}=0$, see \cite{Cresswell-Hogg:2023hdg}.

\begin{figure}[!t]
\includegraphics[width=.9\linewidth]{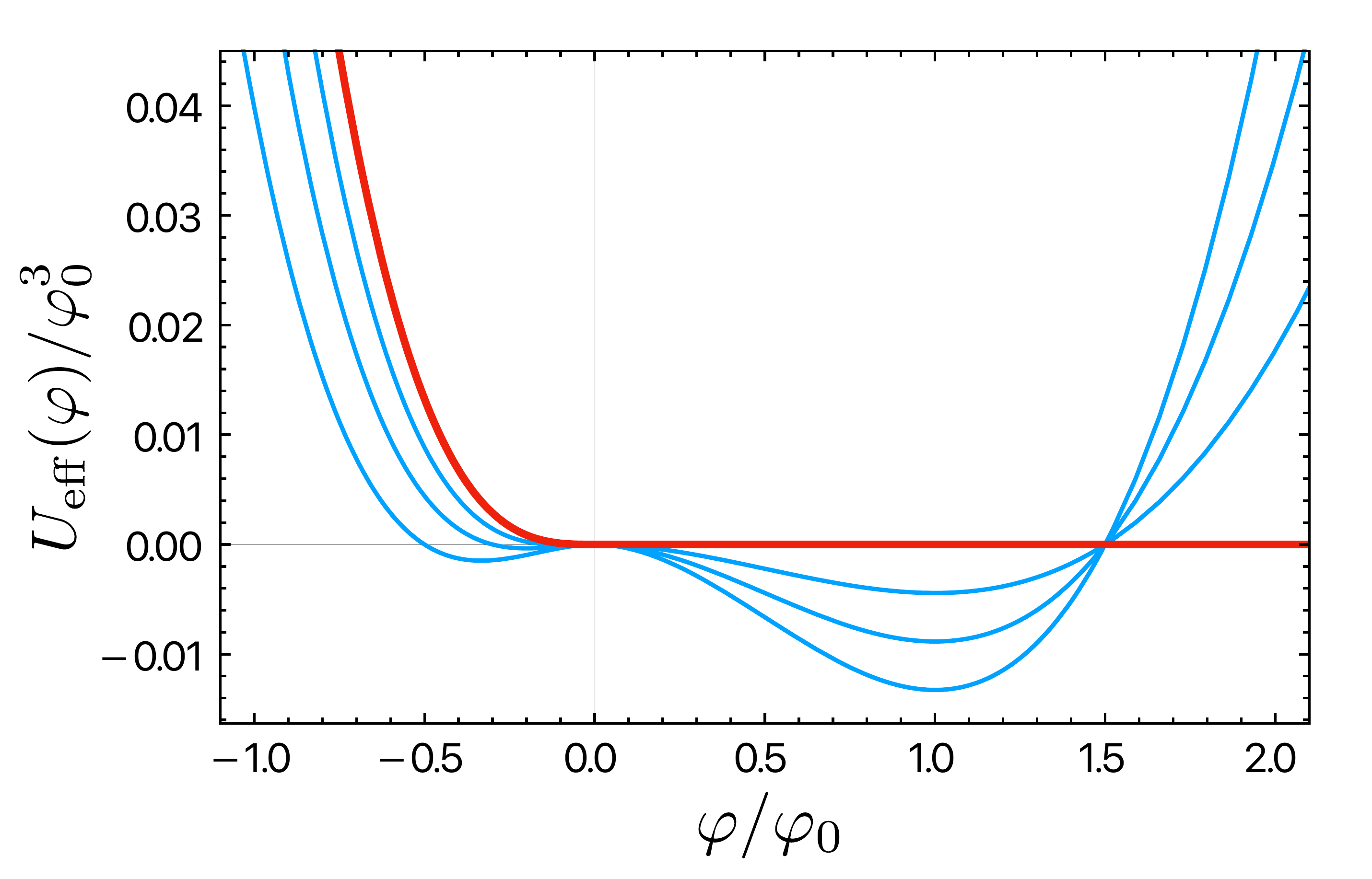}
\caption{Quantum effective potentials with $\varphi_0\neq 0$ (light blue) approaching the   critical endpoint with  broken scale symmetry (red) along  the double-scaling limit as in Fig.~\ref{fig:doubleScaling}.}
\label{fig:UeffDS}
\end{figure}

Next, we consider the  endpoint of the conformal window, $\delta\lam{3}=0$. Its most distinctive feature, the spontaneous appearance of a non-zero $\varphi_0$, can be understood  from a double-scaling limit  sketched  in Fig.~\ref{fig:doubleScaling} (blue line).
Concretely, we start with  non-conformal theories in the massive phase by fixing $\varphi_0>0$ and $\de\lam{2}<0$, and away from the critical endpoint $\de\lam{3}>0$ as in  \eqref{eq:VEV}. 
Performing  the double-scaling limit 
\begin{equation}
\label{eq:DSlimit}
\de\lam{2} \to 0^- \, , \ \de\lam{3} \to 0^+  \, , \, \, \vphiz \text{ fixed}\,,
\end{equation}
we observe that the quantum effective potentials, given by  \eqref{eq:Ueff23}, smoothly approach the half-sidedly flat one at the endpoint of the conformal window, see Fig.~\ref{fig:UeffDS}. 
Since $\vphiz$ has been chosen arbitrarily, the double-scaling limit confirms that the critical potential admits  {\it any} vacuum expectation value $\varphi_0>0$. 
It is interesting to see from  \eqref{eq:VEV} that it is the {\it simultaneous} vanishing of $\delta\lam{2}$ and $\delta\lam{3}$ that enables  $\varphi_0$ to become a free parameter. 
A  virtue of the double-scaling limit is that the  endpoint can  now be accessed as a continuous limit of well-behaved non-conformal theories, as exploited below.

{\it Goldstone boson of scale symmetry breaking.---} 
The dilaton is the Nambu-Goldstone boson of spontaneously broken scale symmetry. 
Its order parameter, the dilaton decay constant $F_D$, is defined through the  matrix element
\begin{equation}
\label{eq:FDdef}
\matel{0}{T_{\mu\nu}}{D(q)} =  
\frac{\Fdec }{ d-1}  
(m_D^2 \eta_{\mu\nu} - q_\mu q_\nu) \;,
\end{equation}
with $T_{\mu\nu}$ the energy momentum tensor, $m_D$ the dilaton mass, $d$ the space-time dimension, 
and $\state{D}$  a one-dilaton state with $\vev{D(\vec{q})|D(\vec{p})} = (2 \pi)^2 {2}E_D \de^{(2)}(\vec{p}-\vec{q})$. 
The dilaton is massless  if it arises  from the  breaking of an exact scale symmetry. 
A non-zero  dilaton mass can be induced by  sources that break  scale symmetry explicitly.

\begin{figure}[!t]
\includegraphics[width=\linewidth]{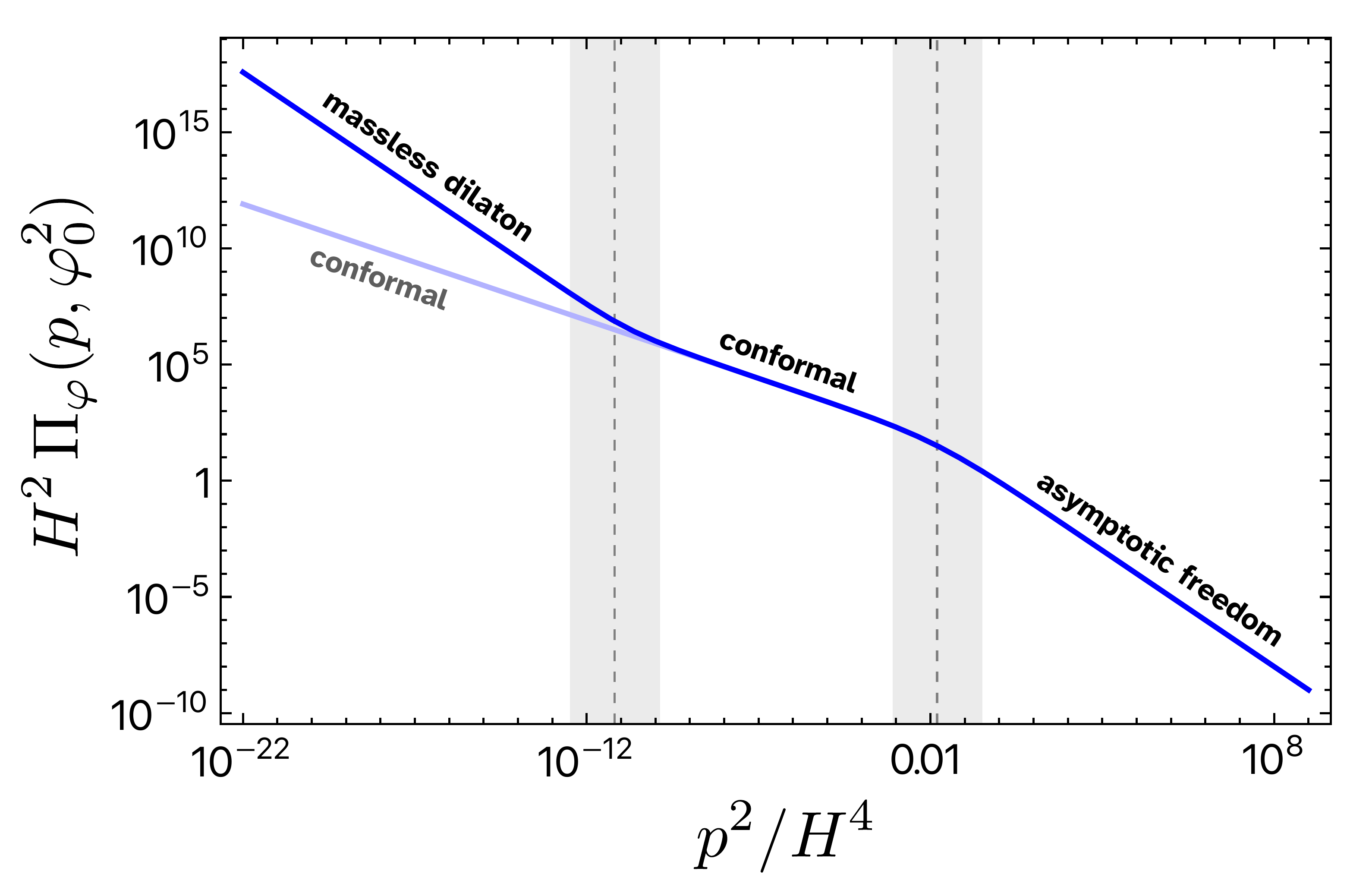}
\caption{Shown are scalar two-point functions $\Pi_\vphi$ from the UV to the IR, comparing theories in the conformal window ($\varphi_0=0$, light blue) to the critical theory with a massless dilaton ($\varphi_0>0$, dark blue); gray bands indicate the scales $\varphi_0$ and $H^2$ ($\vphiz / H^2 = 10^{-6}$).}
\label{fig:2pt}
\end{figure}

\begin{figure}[!t]
\includegraphics[width=.6\linewidth]{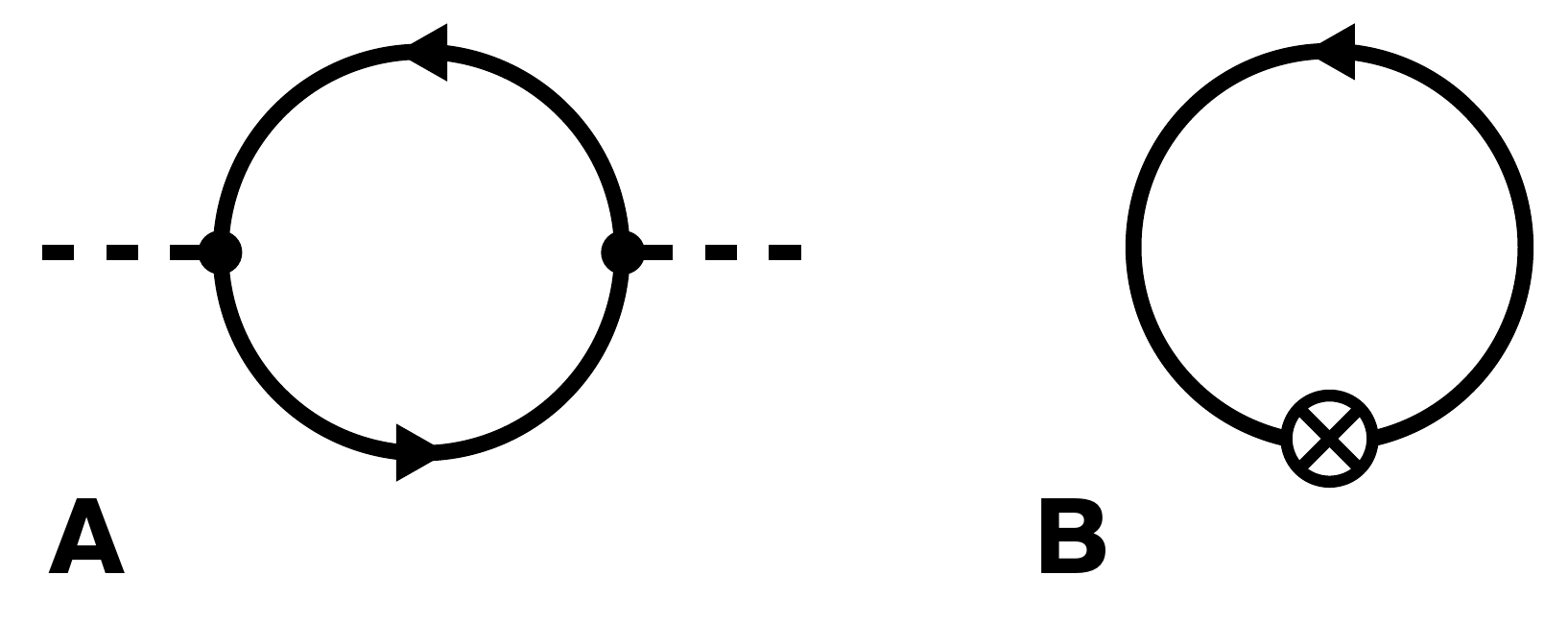}
\caption{Large-$N$ leading quantum corrections to the scalar two-point function (A) and the fermion condensate (B), where the cross stands for an insertion of the operator $\psib\psi$.}
\label{fig:loops}
\end{figure}

{\it The massless dilaton.---} 
The breaking of scale symmetry induces a fermion mass  $m = \vphiz$, and a dilaton becomes  visible in the spectrum as a massless pole in the scalar two-point function 
$\Pi_\vphi ( p, \vphiz ) \equiv \vev{\vphi ( p ) \vphi ( -p )}$  \cite{Semenoff:2024prf}. 
Correlation functions of $\vphi$ can be computed exactly in the large-$N$  limit since boson loops are suppressed.
Integrating out the fermions from the action \eqref{eq:SGNY} yields 
\begin{equation}
\label{eq:Seffvphi}
S_{\rm eff} [ \vphi ] = N \int_x \left\{ \frac{( \partial \vphi )^2}{2 H^2} + \frac{\lambda_3}{3!} \vphi^3 \right\} - N \Tr \ln ( \slashed{\partial} + \vphi ) \, .
\end{equation}
$S_{\rm eff}$ generates all $n$-point functions of $\vphi$ with all large-$N$ leading quantum corrections accounted for via the functional determinant. 
 In particular, the two point function is given by the inverse of  $S_{\rm eff}^{(2)} [ \vphiz ]$ evaluated on the  constant field configuration  minimising the effective potential \eqref{eq:Ueff}. 
In momentum space, one finds
\begin{equation}
\label{eq:phiPropMom}
\Pi_\vphi^{-1} ( p, \vphiz ) = N \Big( \frac{p^2}{H^2} + \lam{3} \, \vphiz + B ( p, \vphiz ) \Big) \; ,
\end{equation}
where $B ( p, \vphiz )$ denotes the one-loop fermion bubble integral with mass $m = \vphiz$ (diagram A in Fig.~\ref{fig:loops}). 
After renormalising the scalar mass  to absorb a momentum-independent UV divergence \footnote{The linear UV divergence of the fermion bubble integral is suppressed in dimensional regularisation. In cutoff regularisation, it entails the  renormalisation of the scalar mass.}, its finite part 
\begin{equation}
\label{eq:bubbleDR}
B ( p, m )  = \frac{1}{4 \pi} \left[ 2 | m | + \frac{p^2 + 4 m^2}{p} \arctan \bigg( \frac{p}{2 | m |} \bigg) \right] 
\end{equation}
expands as $B= \frac{| m |}{\pi} +\frac{p^2}{12\pi|m|}+ {\cal O} ( p^4 )$ for $p^2 \ll m^2$ and
$B= \tfrac p8 +\tfrac {m^2}{2p}+{\cal O}(p^{-2})$ 
for $p^2 \gg m^2$, where $p = \sqrt{p^2}$. Within the conformal window where $\delta\lam{3}>0$ and 
$\vphiz = 0$,  
the two-point function \eqref{eq:phiPropMom} displays a crossover 
from asymptotic freedom $\Pi_\vphi  \sim H^2 / p^2$ in the UV to conformal scaling $\Pi_\vphi  \sim 1 / p$ in the IR. The crossover at the scale $H^2$ also enhances the scaling dimension of $\varphi$ from its canonical dimension  $\De_\vphi = \frac12$  to its non-perturabtive infrared value  $\De_\vphi = 1$ \cite{Cresswell-Hogg:2023hdg,prep} (Fig.~\ref{fig:2pt}, light blue line).

At the endpoint of the conformal window $| \lam{3}| = \lamcrit$ a new mass scale $\vphiz$ appears in addition to $H^2$. In the UV, the  two-point function continues to be governed by asymptotic freedom. If $\vphiz \ll H^2$, we also find a mid-momentum regime with conformal scaling. 
In the IR, however, once $p^2\ll \varphi_0^2$, the fermions effectively decouple. 
Further, the constant  term $\lam{3} \vphi_0<0$ in  \eqref{eq:phiPropMom} 
is exactly cancelled by the leading  term from  the  fermion bubble   \eqref{eq:bubbleDR}, and we are left with 
\begin{equation}\label{eq:expand}
\Pi_\vphi ( p, \vphiz ) = \left( \frac{N}{12 \pi | \vphiz |} + \frac{N}{H^2} \right)^{-1} \frac{1}{p^2} + \dots \,,
\end{equation}
which now shows a massless dilaton pole at vanishing momenta in the spectrum (Fig.~\ref{fig:2pt}, dark blue line). 
It is noteworthy  that the dilaton is truly massless even though  the effects of scale symmetry breaking  only ``emerge'' in the IR limit of the scalar propagator.

{\it The dilaton decay constant.---} 
Using a spectral representation for  correlation functions  \cite{Weinberg:1995mt,Zwicky:2016lka}, the decay constant is extracted from the residue of the dilaton pole 
\begin{equation}
\label{eq:PiRes}
\Pi_\vphi( p, \vphiz ) = \frac{| \matel{0}{\vphi}{D} |^2}{p^2} + \dots \;,
\end{equation}
where the dots stand for higher terms in the spectrum. 
A single soft dilaton theorem  together with $\Delta_\varphi=1$ relate the decay constant  to the residue \cite{Zwicky:2023krx}, giving 
$| \matel{0}{\vphi}{D} |^2 
=  {\vphiz^2}/{F_D^2} +  \ORD(m_D^2) $.
In combination with  
\eqref{eq:expand}, we find 
\begin{equation}
\label{eq:FDresult}
F_D^2 = \frac{N | \vphiz |}{12 \pi} + \frac{N \vphiz^2}{H^2}  \, .
\end{equation}
$F_D^2 \sim  |\varphi_0|$ confirms that $F_D$ is an order parameter 
for  scale symmetry breaking. Note the dependence on the 
the high scale $H^2$,  which drops  out if  $\varphi_0\ll H^2$. 

{\it The fermion condensate.---} The induced fermion mass also triggers a fermion condensate. 
The large-$N$ leading contribution  to the condensate (diagram B in Fig.~\ref{fig:loops}) 
features a short-distance singularity which is regularised via a point-splitting procedure 
$\vev{\bar \psi(0) \psi(x)}   =  - \frac{  N}{2 \pi } \frac{\vphiz}{x}  e^{- \vphiz x}$, and renormalisation is  carried out by applying  
the operator product expansion \cite{Wilson:1969zs,Wilson:1970pq} $\psi(x) \bar \psi(0) = \frac{c_\mathbf{1}}{x} \mathbf{1} + \psi \bar \psi(0) + \ORD(x)$. 
The procedure involves subtracting the identity term, giving the physical condensate as
\begin{equation}
\label{eq:physical}
\vev{\bar \psi \psi}  =  \frac{ N  }{2 \pi} \vphiz^2 \;.
\end{equation}
The condensate is sensitive to the breaking of scale symmetry through the vacuum expectation value $\vphiz$, and vanishes in the limit of unbroken scale symmetry.

\begin{figure}[!t]
\includegraphics[width=\linewidth]{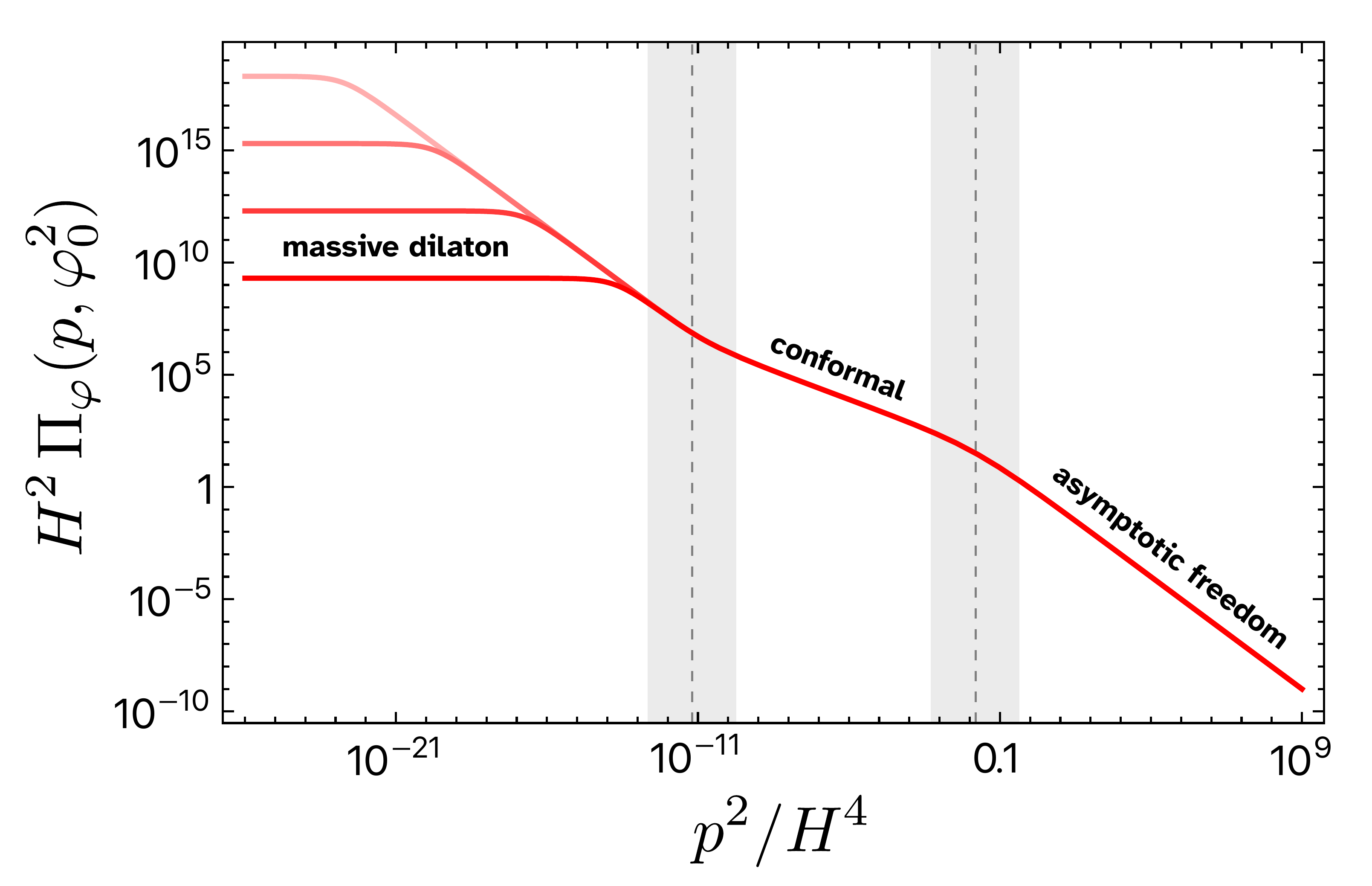}
\caption{Same as Fig.~\ref{fig:2pt}, showing two-point functions $\Pi_\varphi$ close to the critical theory with a gap in the spectrum due to a massive dilaton from scalar mass perturbations; $\delta\lam{2}$ from \eqref{eq:VEV}, $\vphiz / H^2 = 10^{-6}$ and $\de\lam{3} = 10^{-12}, \, 10^{-9} , \, 10^{-6}, \, 10^{-3}$  (top to bottom).}
\label{fig:2ptMassive}
\end{figure}
{\it The massive dilaton.---} 
The dilaton becomes massive in the presence of small perturbations that break scale symmetry explicitly. 
Let us first add a  scalar mass term $\delta \bar\lambda_2 \phi^2$ to the fundamental theory \eqref{eq:SGNY}, which  translates to adding a mass term $\delta\lam{2}\varphi^2$ to the effective potential  as in \eqref{eq:Ueff23}. 
The perturbation contributes a constant term to the inverse propagator \eqref{eq:phiPropMom}, $\Pi_\vphi^{-1} \to \de\lam{2} + \Pi_\vphi^{-1}$, shifting the dilaton pole in \eqref{eq:PiRes} away from vanishing momentum.

To find the induced dilaton mass, we consider non-conformal theories with effective potentials \eqref{eq:Ueff23} and $\vphiz \geq 0$, and perform the double-scaling limit \eqref{eq:VEV}, \eqref{eq:DSlimit} to approach the critical endpoint. 
To linear order, we find
\begin{equation}
\label{eq:mDscalar}
m_D^2 = \frac{6 \pi H^2 \vphiz^2}{H^2 + 12 \pi \vphiz} \, \de\lam{3} \,.
\end{equation}
Even though the dilaton is an IR effect, its induced mass is sensitive to  both fundamental scales   $H^2$ and  $\varphi_0$
due to \eqref{eq:expand}. 
In terms of $\delta\lam{2}<0$ we find $m_D^2 = \frac{-12 \pi H^2 \vphiz}{H^2 + 12 \pi \vphiz} \, \de\lam{2}$. 
Fig.~\ref{fig:2ptMassive} shows   two-point functions  $\Pi_\varphi$ in the presence of scalar mass perturbations  assuming  $m_D^2 \ll \vphiz \ll H^2$. In  the large- and mid-momentum regime, results are  indistinguishable from the  theory with a massless dilaton, see Fig.~\ref{fig:2pt}. 
At low momenta, however, the propagator is gapped below the induced  mass \eqref{eq:mDscalar}.

Next, we consider  a fermion mass perturbation  $\de m \mkern1mu \psib\psi$ with quantum effective potential
\begin{equation}\label{eq:Ueffm}
U_{\rm eff} ( \vphi ) = \frac{1}{3!} \de\lam{3} \, \vphi^3 + \frac{1}{3!} \lamcrit ( | \vphi + \de m |^3 - \vphi^3 ) \, .
\end{equation}
The non-conformal theories  \eqref{eq:Ueffm} are globally stable with a 
minimum at $\varphi_0>0$ provided  $\delta m<0$ and
\begin{equation}
\label{eq:dl3fermion}
\de\lam{3} = -2 \lamcrit \, \frac{\de m}{\vphiz} + {\cal O} ( \de m^2 ) \, .
\end{equation}
Then, in analogy to \eqref{eq:DSlimit}, a  double-scaling limit approaching the critical theory 
is given by $\de m \to 0^-$ and $\de\lam{3} \to 0^+$ with $\varphi_0$ held fixed by \eqref{eq:dl3fermion}. 
The fermion mass  modifies the inverse  propagator \eqref{eq:phiPropMom}, $B ( p, \vphiz )\to B ( p, \vphiz + \de m )$, and shifts the dilaton pole  in \eqref{eq:PiRes}  to a non-zero value. The dilaton mass to linear order in $\delta\lam{3}$ continues to be  given by   \eqref{eq:mDscalar}.   
Expressed in terms of $\delta m<0$ using \eqref{eq:dl3fermion}, it reads $m_D^2 = \frac{-12  H^2 \vphiz}{H^2 + 12 \pi \vphiz} \, \delta m$. 
We observe that the result for the induced dilaton mass \eqref{eq:mDscalar} is independent of the  mechanism for explicit symmetry breaking. 

Combining  the decay constant \eqref{eq:FDresult} with the induced mass \eqref{eq:mDscalar}, 
we find that any dependence on the UV scale disappears in their product to linear order in  $\delta\lam{3}$  
\begin{equation}
\label{eq:mDFD}
m_D^2 F_D^2 = \tfrac12 N \vphiz^3 \, \de\lam{3} \,.
\end{equation}
The reason for this is that $m_D^2 F_D^2/\varphi_0^2$ is proportional to the constant term generated in $\Pi^{-1}_\varphi$ 
by explicit symmetry breaking, which is $H^2$-independent.

{\it  Consistency with  soft dilaton theorems.---} 
It is useful to confirm \eqref{eq:mDFD} using double-soft dilaton theorems \cite{Zwicky:2023bzk,Zwicky:2023krx} and the scaling dimensions of underlying perturbations \cite{Cresswell-Hogg:2023hdg,prep}. 
This line of reasoning employs the Feynman-Hellmann  theorem, e.g.  \cite{Zwicky:2023bzk}, to relate the dilaton mass $m_D^2 =  \de \la\, \matel{D}{ \Op}{D}  + \ORD(\de \la^2)$
to a perturbation of the Hamiltonian $\de H = \de \la\, \Op$. Together with  the recently derived soft theorem  
$\matel{D}{\Op}{D} = \De_{\Op}(\De_{\Op}-d) \vev{\Op} / F_D^2$  \cite{Zwicky:2023krx}, this leads to 
\begin{equation}
\label{eq:soft}
m_D^2 F_D^2 =  \De_{\Op}(\De_{\Op}-d) \,\de \la\, \vev{\Op} \;,
\end{equation}
a relation expressing $m_D^2 F_D^2 $ in terms of the scaling dimension and the vacuum expectation value of the  perturbation $\Op$. 

To compare with \eqref{eq:mDFD}, we use  perturbations of the form 
$\de  H =  \de m \,\Oqq +N(  \frac{\de \la_2 }{2!} \varphi^2 + \frac{\de \la_3 }{3!} \varphi^3)$, 
with scaling dimensions $\De_{\vphi^n} = n$ and $\De_{\Oqq} =1$. 
For a scalar mass perturbation  
\eqref{eq:VEV}, \eqref{eq:DSlimit} with $\de m =0$ we obtain 
\begin{equation}
\label{eq:GMORscalar}
m_D^2 F_D^2 =  -N\, \de \la_2\, \vev{\vphi^2} \,, 
\end{equation} 
which is in exact agreement with \eqref{eq:mDFD} once rewritten in terms of $\de\lam{3}$ from  \eqref{eq:VEV}, and  using the large-$N$ factorisation of the vacuum expectation value $\vev{\vphi^2} = \vphiz^2$.
For a fermion mass perturbation \eqref{eq:dl3fermion} with $\de \la_2 =0$, we find
\begin{equation}
\label{eq:GMORfermion}
m_D^2 F_D^2 =  -2\, \de m\, \vev{\psib\psi} \,,
\end{equation} 
which equally agrees with   \eqref{eq:mDFD} once expressed in terms of \eqref{eq:dl3fermion} and  substituting the fermion condensate \eqref{eq:physical}.  
As an additional bonus this serves as a consistency check of the fermion condensate per se. 
Notice that the  cubic term $\sim \varphi^3$, added to ensure stability of the perturbed theory, 
does not contribute to \eqref{eq:GMORscalar},  \eqref{eq:GMORfermion} because its exact marginality  implies that the  factor $(\De_{\vphi^3} -3)$ in 
\eqref{eq:soft} vanishes identically \footnote{In the case where a \emph{single} operator contributes to $\TEMT=\Op$, for the dilaton mass and the decay constant, its scaling dimension must be $\De_{\Op}=d-2$ \cite{Zwicky:2023krx}. Here, instead, \emph{several} operators are involved, which can lead to different scaling dimensions.}.

{\it Summary, discussion, and outlook.---} 
We have studied spontaneous scale symmetry breaking and the dilaton in an asymptotically free Gross-Neveu-Yukawa theory.  
Quantum effects generate an exactly marginal   operator that controls a finite line of  
fixed points with spontaneously broken scale symmetry and a  dilaton at its endpoint. 
Vacuum stability imposes bounds on the conformal window  (Fig.~\ref{fig:UeffCrit}), and the  scalar propagator exhibits  scaling regimes with and without  a massless (Fig.~\ref{fig:2pt}) or massive (Fig.~\ref{fig:2ptMassive})  dilaton.  
Our model provides an explicit example of a massless dilaton compatible with 
a renormalisation group flow and a trace anomaly \cite{prep}. 

To access the dilaton, we have advocated a new technique dubbed double-scaling limit (Figs.~\ref{fig:doubleScaling} and~\ref{fig:UeffDS}). 
It proved most useful to find the result \eqref{eq:mDFD}, and a universal form for the induced dilaton mass \eqref{eq:mDscalar}, irrespective of the symmetry breaking mechanism. 
In this context, it is worth noting that \eqref{eq:mDFD} is only sensitive to the scale $\varphi_0$, whereas quantities such as the decay constant, induced mass, or the scalar propagator, in general depend on  {\it both}   fundamental scales, $H^2$  and $\varphi_0$. 

The computation of low-energy parameters in terms of the microscopic theory, but with different methods, provides  useful consistency  checks. Here, we have used   soft  dilaton theorems \cite{Zwicky:2023krx}  to independently  determine $m_D^2F_D^2$, see 
\eqref{eq:soft}. The link with  the result  \eqref{eq:mDFD} can  now be exploited in new ways, for example to  find   operator scaling dimensions  
 from small perturbations in a double-scaling limit, or vice versa  \cite{prep}.

In the limit  where  the high scale   is removed $1/H^2\to 0$, a series of  CFTs  emerges from our models inside the conformal window. 
Then, the endpoint offers a rare example of  a non-supersymmetric CFT with an exact vacuum moduli space parametrised by a nonzero 
$\varphi_0$, that may  guide general studies of  {CFT$+\varphi_0$}  theories \cite{Karananas:2017zrg,Zwicky:2023krx,Cuomo:2024vfk}. 
Looking beyond, our model also offers insights into CFT$+\varphi_0$ theories under 
conformal perturbations, e.g.~\cite{Gaberdiel:2008fn,Komargodski:2016auf}, and it would be instructive to extract  results such as \eqref{eq:mDFD} from this perspective. 

Finally, it would  be interesting to extend our methods to  gauge theories with matter where conformal 
windows arise abundantly, or even QCD, where it has  been speculated that a dilaton could be 
present in or below the conformal window. 
Under this hypothesis \cite{Zwicky:2023krx}, some features are already shared with our template. 
For instance, the presence of an exactly marginal operator, integer scaling dimensions and a 
fermion mass anomalous dimension of unity,  make this topic  worthy of further study and exploration.\\

{\it Acknowledgements.---} This work has  been  supported by the {\it Science and Technology Facilities Council} (STFC) under the Consolidated Grants T/X000796/1 (CCH, DFL) and  ST/P0000630/1 (RZ), and by a CERN Associateship (DFL).

\bibliography{Dil-refs}
\bibliographystyle{mystyle2}

\end{document}